\documentclass[dvipdfmx,conference]{IEEEtran}
\usepackage{graphicx}
\usepackage{latexsym}
\usepackage{url}
\usepackage{amsmath}
\usepackage{amsthm}
\usepackage{amssymb}
\usepackage{bbm}
\usepackage{graphicx}
\usepackage{booktabs}
\usepackage{float}

\newcommand{\nup}{\nu^\perp}
\newcommand{\mup}{\mu^\perp}

\newcommand{\dg}{d_\mathrm{g}}
\newcommand{\dl}{d_\mathrm{l}}

\newcommand{\dr}{d_\mathrm{r}}

\newcommand{\dgp}{d_\mathrm{g}^\perp}
\newcommand{\dlp}{d_\mathrm{l}^\perp}

\newcommand{\drp}{d_\mathrm{r}^\perp}
\newtheorem{theorem}{Theorem}
\newtheorem{discussion}{Discussion}
\newtheorem{definition}{Definition}

\newtheorem{lemma}{Lemma}

\newcommand{\Rpre}{R_\mathrm{pre}}
\newcommand{\Rtotal}{R_\mathrm{total}}

\newcommand{\xp}{\x^{\perp}}

\newcommand{\x}{\bm{x}}
\newcommand{\y}{\bm{y}}

\newcommand{\D}{\bm{D}}
\newcommand{\f}{\bm{f}}
\newcommand{\g}{\bm{g}}

\newcommand{\hC}{\widehat{\mathcal{C}}}
\newcommand{\C}{\mathcal{C}}
\newcommand{\Cp}{\mathcal{C}^\perp}
\newcommand{\Xc}{\mathcal{X}}

\newcommand{\Ec}{\mathcal{E}}

\newcommand{\ehs}{\epsilon^{\mathrm{Sha}}}

\newcommand{\es}{\epsilon_{\scalebox{0.5}{}}^{\scalebox{0.5}{\rm{Sha}}}}

\newcommand{\xom}{x_{1}^\perp}
\newcommand{\xtm}{x_{2}^\perp}
\newcommand{\xoh}{x_1}
\newcommand{\xth}{x_2}

\newcommand{\xmn}{\bm{x}^\perp}
\newcommand{\xha}{\bm{x}}

\newcommand{\Umn}{U^\perp}
\newcommand{\Uha}{U}

 \usepackage{color} 
\usepackage{amsmath}
\usepackage{amsthm}
\usepackage{amssymb}
\usepackage{bm}
\usepackage{graphicx}
\usepackage{booktabs}
\usepackage{float}

\usepackage{mathtools}
\mathtoolsset{showonlyrefs}

\ifCLASSINFOpdf
\else
\fi
\begin{document}
%
\title{Spatially-Coupled Precoded Rateless Codes with Bounded Degree Achieve the Capacity of BEC under BP decoding}
\author{
\IEEEauthorblockN{Kosuke Sakata,  Kenta Kasai and Kohichi Sakaniwa}
\IEEEauthorblockA{Department of Communications and 
Computer Engineering,\\
 Tokyo Institution of Technology\\
Email: \{sakata,kenta,sakaniwa\}@comm.ce.titech.ac.jp}

}
\maketitle

\begin{abstract}
Raptor codes are known as precoded rateless codes that achieve the capacity of BEC. 
However the maximum degree of Raptor codes needs to be unbounded to achieve the capacity. 
In this paper we prove that spatially-coupled precoded rateless codes achieve the capacity with bounded degree under BP decoding. 
\end{abstract}
\IEEEpeerreviewmaketitle

 \section{Introduction}
Spatially-coupled (SC) low-density parity-check (LDPC) codes attract much attention due to 
their capacity-achieving performance under low-latency memory-efficient sliding-window belief propagation (BP) decoding.
The studies on SC-LDPC codes date back to the invention of convolutional LDPC codes by Felstr{\"o}m and Zigangirov \cite{zigangirov99}. 
Lentmaier {\itshape et al.}! observed that the BP threshold of regular SC-LDPC codes coincides with the maximum a posterior (MAP) threshold of the underlying block LDPC 
codes with a lot of accuracy  by density evolution  \cite{lentmaier_II}. 
Kudekar {\it et al.}~ proved that SC-LDPC codes achieve the MAP threshold of BEC \cite{5695130} 
and the binary-input memoryless output-symmetric (BMS) channels \cite{6589171} under BP decoding. 

Rateless codes are a class of erasure-recovering codes which produce limitless sequence of encoded bits 
from $k$ information bits so that receivers can recover the $k$ information bits from arbitrary $(1+\alpha)k/(1-\epsilon)$ received symbols from BEC($\epsilon$). 
We denote  {\itshape overhead} by $\alpha$. 
Designing rateless codes with vanishing overhead is desirable, which implies the codes achieve the capacity of BEC($\epsilon$). 
LT codes \cite{1181950} and  raptor codes \cite{raptor} are rateless codes that
achieve vanishing overhead $\alpha\to 0$ in the limit of large information size over the BEC.
By a nice analogy between the BEC and the packet erasure channel \textcolor{black}{(e.g., Internet)}, 
rateless codes have been successfully adopted by several industry standards. 

A raptor code can be viewed as concatenation of an outer high-rate LDPC code and infinitely many single parity-check codes of length $d$, 
where $d$ is chosen randomly with probability $\Omega_d$ for $d\ge 1$. 
Raptor codes need to have unbounded maximum degree $d$ for $\Omega_d\neq 0$. 
This leads to a computation complexity problem both at encoders and decoders. 

Aref and Urbanke \cite{ITW_AREF_URBANKE} investigated the potential advantage of universal achieving-capacity property of 
SC low-density generator matrix (LDGM) codes. 
They observed that the decoding error probability steeply decreases with overhead $\alpha=0$ with bounded maximum degree over various BMS channels. 
However the decoding error probability was proved to be bounded away from 0 with bounded maximum degree for any $\alpha$. 
This is explained from the fact that there are a constant fraction of bit nodes of degree 0. 

The authors presented empirical results in \cite{HSU_MN_IEICE}  showing 
that SC MacKay-Neal (MN) codes and SC Hsu-Anastasopoulos (HA) codes achieve the capacity of BEC with bounded maximum degree. 
Recently a proof for SC-MN codes are given in \cite{ISIT_OJKP} by using potential functions. 
It was observed that the SC-MN codes and SC-HA codes have the BP threshold close to the Shannon limit in \cite{6400949} over BMS channels.

In \cite{6620664}, the authors proposed SC precoded rateless codes by concatenating regular SC-LDPC codes and regular SC-LDGM codes. 
We derived a lower bound of the asymptotic overhead from stability analysis for successful decoding by density evolution. 
We observe that with a sufficiently large number of information bits, the asymptotic overhead and the decoding error rate approach 0 with bounded maximum degree.

In this paper we give a proof that the SC precoded rateless codes with possible smallest degree achieve the capacity of BEC under BP decoding.
The proof needs to use duality property of potential functions, since this avoid facing difficulties to prove an inequality with a parametrically described constraint. 
 \section{Preliminaries}

\subsection{$(\dl,\dr,\dg)$ Precoded-Rateless Code $\C$}
Throughout this paper, we deal with a randomly constructed code as a random variable. 
In this section, we define a precoded-rateless code $\C$. 
This induces spatially-coupled  precoded-rateless code  $\widehat{\C}$. 
Let $k$ be the number of information bits. For $\dl\ge 2,\dr\ge 3,\dg\ge 2$, we define $(\dl,\dr,\dg)$  precoded-rateless code  as follows. 
Construction of $(\dl,\dr,\dg)$ precoded-rateless code  $\C$ involves two encoding steps: outer code (precode) and inner code.  
First $k$ information bits are encoded by $(\dl,\dr)$ regular LDPC codes into $M$ bits $x(0),\dotsc,x(M-1)$. 
The coding rate of $(\dl,\dr)$ regular LDPC codes $\Rpre$ is given as follows. 
\begin{align*}
 \Rpre=1-\frac{\dl}{\dr}.
\end{align*}
From this, we have $k=\Rpre M$. 

We further pre-coded $M$ into infinitely long code bits by inner code. 
At each time $t\in [1,\infty)$, repeat 1 and 2. 
\begin{enumerate}
 \item Choose $\dg$ indices $l^{(t)}_1,\dotsc,l^{(t)}_{\dg}\in[0,M-1]$ uniformly at random. 
 \item Calculate the sum of the following $\dg$ bits and transmit the sum. 
      \begin{align*}
x(l^{(t)}_{1})+\cdots+x(l^{(t)}_{\dg}).
\end{align*}
\end{enumerate}

Consider a receiver receives $n$ symbols $y^{(1)},\dotsc,y^{(n)}$ through BEC$(\epsilon)$. 
We define the {\itshape overhead} as 
\begin{align}
 \alpha=\frac{n}{k}(1-\epsilon)-1.\label{overhead}
\end{align}
Zero overhead $\alpha=0$ i.e. $\frac{k}{n}:=(1-\epsilon)$ implies capacity achieving over BEC$(\epsilon)$. 

The factor graph \cite{910572} for decoding consists of $M$ bit nodes and $\frac{\dl}{\dr}M$ parity-check nodes of pre-code and $n$ channel factor nodes 
$\mathbbm{1}[x(l^{(t)}_{1})+\cdots+x(l^{(t)}_{\dg})=y^{(t)}]$ for $t=1,\dotsc,n$.
Let $\Lambda_d$ be the probability that a bit node connects to $d$ channel nodes and let $\beta$ be the expected number of channel nodes to which a bit node connect to. 
For $M\to \infty$, we have 
\begin{align*}
 &\beta:=\frac{\dg \Rpre(1+\alpha)}{1-\epsilon}, \\
 &\Lambda(x)=\sum_{d\ge 0}\Lambda_dx^d =e^{-\beta(1-x)}=\sum_{d\geq0}\frac{{\beta}^de^{-{\beta}}}{d!}x^d,\\
 &\Rtotal:=\frac{k}{n}=\frac{\dg}{\beta}\Rpre=\frac{\dg}{\beta}\Bigl(1-\frac{\dl}{\dr}\Bigr), 
\end{align*}
where $\Rtotal$ is equal to the coding rate. 
Assume the transmission takes place over BEC($\epsilon$), Shannon limit is given by $\ehs:=1-\Rtotal=1-\frac{\dg}{\beta}\Bigl(1-\frac{\dl}{\dr}\Bigr)$.

The multivariate degree distribution (dd) \cite{met,ISIT_OJKP} of $(\dl,\dr,\dg)$ precoded-rateless code is given by
\begin{align}
 \nu(\x;\epsilon) = \frac{\dg}{\beta}x_1^{\dl}\Lambda(x_2),\  \mu(\x;\epsilon)=\frac{\dg \dl}{\beta \dr}x_1^{\dr} + \epsilon x_2^{\dg}.
\end{align}
\subsection{Vector Admissible System}
\begin{definition}
\label{def:vas}
For $\Xc\triangleq[0,1]^d$, and functions
 $F : \Xc \times [0, 1]\to \mathbb{R}$, $G :\Xc \times [0, 1]\to \mathbb{R}$ satisfying $G(\bm{0},\epsilon)=0$ and a $d \times d$ positive diagonal matrix $\D$, 
consider the following recursion
  \begin{align}
    \label{eq:1}
    \x^{(0)} &= \bm{1},\\\
    \x^{(\ell+1)} &= \f(\g(\x^{(\ell)};\epsilon);\epsilon), 
  \end{align}
where we define $\f : \Xc \times [0,1] \to \Xc$, $\g: \Xc \times [0, 1]\to \mathbb{R}$, $F'(\x;\epsilon) = \f(\x;\epsilon) \D$ and 
$G'(\x;\epsilon) = \g(\x;\epsilon) \D$. 
We say that  the system $\x^{(\ell)}_{\ell\ge 0}$ or equivalently $(\f,\g)$ is a vector admissible system if
\begin{enumerate}
\item $\f,\g$ are twice continuously differentiable,
\item $\f(\x;\epsilon),\g(\x;\epsilon)$ are non-decreasing (w.r.t. $\preceq$) in $\x$ and $\epsilon$,
\item $\f(\g(\bm{0};\epsilon);\epsilon)=\bm{0}$, and $F(\g(\bm{0};\epsilon);\epsilon)=0$.
\end{enumerate}
\end{definition}

\begin{definition}[{\cite[Def. 2]{simple_proof_vector_turbo}}]
  \label{def:potential}
Generally, the potential function $U(\x;\epsilon)$ of a vector admissible system $(\f ,\g)$ is defined as follows. 
 \begin{align}
   U(\x ;\epsilon) 
      &=\g(\x;\epsilon)\D\x^T-G(\x;\epsilon)-F(\g(\x;\epsilon);\epsilon)\label{U}.
    \end{align}
\end{definition}
For a code with $d$-variable dd  $(\mu,\nu)$ is given by 
\begin{align}
 y_{j}^{(\ell)} & =
1-\frac{\mu_{j}\left(\bm{1}-\bm{x}^{(\ell)};1-\epsilon\right)}{\mu_{j}(\bm{1};1)} =:\left[\g\left(\x^{(\ell)};\epsilon\right)\right]_{j},\label{044746_25Jan14}\\
 x_{j}^{(\ell+1)} & =
\frac{\nu_{j}\left(\y^{(\ell)};\epsilon\right)}{\nu_{j}(\bm{1};1)}=: \left[\f\left(\y^{(\ell)};\epsilon\right)\right]_{j},\label{044809_25Jan14} 
\end{align}
for $i=1,\dotsc,d$, where $\nu_{j}(\x;\epsilon)\triangleq\frac{\partial}{\partial x_{j}}\nu(\x;\epsilon)$ and  $\mu_{j}(\x;\epsilon)\triangleq\frac{\partial}{\partial x_{j}}\mu(\x;\epsilon)$.
Such a system $(\f,\g)$ is a vector admissible system and the potential function  is given as follows \cite{ISIT_OJKP}鐚
\begin{align}\label{poten}
U(\x;\epsilon)&=\mu(\bm{1},1-\epsilon)-\mu(\bm{1}-\x;1-\epsilon)-\nu(\g(\x;\epsilon);\epsilon)\nonumber\\
& -\sum_{j=1}^{d}x_{j}\mu_{j}\left(\bm{1}-\bm{x};1-\epsilon\right).
\end{align}

\begin{definition}[{\cite[Def. 7]{simple_proof_vector_turbo}}]
\label{def:PotentialThreshold}
  The potential threshold is 
  \begin{equation}
    \label{eq:3}
    \epsilon^{*} \triangleq \sup \{ \epsilon \in \Ec \mid \inf\nolimits_{\x \in \mathcal{F}(\epsilon)} U(\x;\epsilon) > 0 \}.
  \end{equation}
  This is well defined only if $\{ \epsilon \in \Ec \mid \inf\nolimits_{\x \in \mathcal{F}(\epsilon)} U(\x;\epsilon) > 0 \} \neq \emptyset$.
  Let $\epsilon_s^*$ be the uncoupled system threshold defined in \cite[Def. 6]{simple_proof_vector_turbo}.
  For $\epsilon_s^* < \epsilon < \epsilon^*$, the \emph{energy gap},
  $$\Delta E(\epsilon) \triangleq \max_{\epsilon'\in [\epsilon,1]} \inf_{ \x\in \mathcal{F}(\epsilon')} U(\x;\epsilon'),$$
  is well defined and  strictly positive.
\end{definition}

\begin{definition}[{\cite[Def.~9]{simple_proof_vector_turbo}}]
\label{def:SpatiallyCoupled} 
We define spatially-coupled system $\{\x_i^{(\ell)}\}_{i\ge 0,\ell\ge 0}$ of coupling number $L$ and coupling width $w$ is defined as
 \begin{align}
  \x^{(0)}_i &=1,\\
  \x^{(t+1)}_i &=\frac{1}{w}\sum_{k=0}^{w-1}\f\Bigg(\frac{1}{w}\sum_{j=0}^{w-1}\g(\x^{(t)}_{i+j-k};\epsilon_{i+j-k});\epsilon_{i-k}\Bigg),\label{054448_25Jan14}\\
  \epsilon_{i} &=
  \begin{cases}
   \epsilon& i\in \{0,\dotsc,L-1\}\\
   0& i\notin \{0,\dotsc,L-1\}.
  \end{cases}
 \end{align}
\end{definition}
If we take $(\f,\g)$ as \eqref{044746_25Jan14} and \eqref{044809_25Jan14}, SC-system $\{\x_i^{(\ell)}\}_{i\ge 0,\ell\ge 0}$ gives 
the DE of spatially-coupled code with dd $(\mu,\nu)$. 

\begin{theorem}[{\cite[Thm.~1]{simple_proof_vector_turbo}}]
\label{thm:sc_theorem} 
For a vector admissible system~$(\f,\g)$
with potential threshold $\epsilon^*$. 
There exists a constant $K_{\f,\g}$ such that 
if $\epsilon < \epsilon^*$ and $w > (d K_{\f,\g})/( 2\Delta
E(\epsilon))$, the only fixed point of the SC-system
of coupling width $w$ is $\bm{0}$.
\end{theorem}

From this theorem, it follows that  a potential threshold of DE system \eqref{044746_25Jan14} and \eqref{044809_25Jan14}, 
can be achieved on SC-system in \eqref{054448_25Jan14} by BP decoding. 

\subsection{Dual Code}
For a code $\C$ with dd $(\nu, \mu)$, we define the dual code $\Cp$ as a code with 
dd $(\nu^\perp, \mu^\perp)=(\mu, \nu)$. 
The DE of the dual code is given \cite{ISIT_OJKP} by $\x^{(\ell+1)}=\f^{\perp}(\g^{\perp}(\x^{(\ell)};\epsilon);\epsilon)$, where 
\begin {align*}
 \f^\perp(\x,\epsilon)&\triangleq \bm{1}-\g(\bm{1}-\x;1-\epsilon),\\
 \g^\perp(\y,\epsilon)&\triangleq \bm{1}-\f(\bm{1}-\y;1-\epsilon).
\end{align*}

Next lemma gives duality property over $\C$ and $\Cp$ \cite{ISIT_OJKP}.
\begin{lemma}[{\cite[Lem.~2]{ISIT_OJKP}}]
\label{lem:duality} 
Let $\C$ be a code with dd $(\nu,\mu)$. 
Let $(\xp,\epsilon)$ be a fixed point of the DE of $\Cp$. In precise, $\xp=\f^\perp(\g^\perp(\xp;\epsilon))$. 
Then, $(\x,1-\epsilon)$ is a fixed point of DE of $\C$, where $\x\triangleq\f(\bm{1}-\xp;1-\epsilon)$. 
Let the potential functions of $\C$ and $\C^\perp$ be $U$ and $U^\perp$, respectively. 
Then it holds that 
\begin{align*}
U^\perp(\xp;\epsilon)=U(\f(\bm{1}-\xp;1-\epsilon);1-\epsilon)\\+\nu(\bm{1},1-\epsilon)-\mu(\bm{1},\epsilon).
\end{align*}
\end{lemma}
From this lemma, by switching $\C$ and $\Cp$, it follows that 
$(\x^\perp,1-\epsilon)$ is a fixed point of DE of $\C^\perp$, where $\xp\triangleq\f^{\perp}(\bm{1}-\x;1-\epsilon)$ 
and 
\begin{align}
\begin{split}
 U(\x;\epsilon)=U^\perp(\f^\perp(\bm{1}-\x;1-\epsilon);1-\epsilon)\\+\mu(\bm{1},1-\epsilon)-\nu(\bm{1},\epsilon).
\end{split}\label{073849_25Jan14}
\end{align}

\section{Proof of Capacity Achievability}
In this section, we prove that the SC $(\dl,\dr,\dg)$ precoded-rateless code $\hC$ achieves the capacity of BEC under BP decoding for some bounded $(\dl,\dr,\dg)$. 
It is sufficient to show that the potential function of DE system of $\C$ is equal to Shannon threshold of $\C$.
The strategy of the proof is as follows. 
We first prove that the potential threshold of DE system of $\Cp$ is equal to Shannon threshold of $\Cp$.
Then we prove, by using Lemma \ref{lem:duality}, that the potential threshold of DE system of $\C$ is equal to Shannon threshold of $\C$.

Define $\dlp=\dr, \drp=\dl,\dgp=\dg$. 
The dd of the dual code of $\C$ is given as follows. 
\begin{align}
\label{MN:dist}
\begin{split}
 &\nup(\x;\epsilon) =\frac{\dgp \drp}{\beta \dlp}x_1^{\dlp} + \epsilon x_2^{\dgp}, \\
 &\mup(\x;\epsilon) =\frac{\dgp}{\beta}x_1^{\drp}\Lambda(x_2).
\end{split}
\end{align}
\begin{discussion}
 Some readers might think that it is more natural to prove directly that the potential threshold of $\C$ is equal to Shannon threshold of $\C$. 
 If we choose that proof, we have to prove an inequality with a parametrically described constraint. 
 The proof via $\Cp$ avoids such a difficult problem. 
 Furthermore, some readers might think that why we do not use spatially-coupled dual code $\Cp$, i.e., $\widehat{\Cp}$ as a rateless code.
 Unfortunately and interestingly, $\widehat{\Cp}$ can not be used as a rateless codes since the inner code can not be viewed as an LDGM code. 
\end{discussion}
\subsection{Potential Function of $\Cp$ at Fixed Points}
From \eqref{poten} and \eqref{MN:dist},  it follows that the  DE system of $\Cp$ is given by 
\begin{align}
\label{065644_25Jan14}
\begin{split}
 x_{1}^{(\ell+1)}&=(1-(1-x^{(\ell)}_{1})^{\drp-1} \Lambda(1-x^{(\ell)}_{2}))^{\dlp-1},\\
 x_{2}^{(\ell+1)}&=\epsilon(1-(1-x^{(\ell)}_{1})^{\drp}\Lambda(1-x^{(\ell)}_{2}))^{\dgp-1}.
\end{split}
\end{align}
The potential function of the DE system is given as follows. 
\begin{align}\label{po-mn}
 &U^\perp(x_1,x_2,\epsilon)=\frac{\dgp}{\beta}-\Bigl( \frac{\dgp}{\beta}(1-x_1)^{\drp}\Lambda(1-x_2) \Bigr)\nonumber\\
 &\quad-\Bigl(\frac{\dgp \drp}{\beta \dlp} x_1^{\frac{\drp}{\dlp-1}}+\epsilon(1-(1-x_1)(1-x_1^{\frac{1}{\dlp-1}}) )^{\dgp} \Bigr)\nonumber\\
 &\quad-\Bigl(\frac{\dgp \drp}{\beta}x_{1}(1-x_{1})^{\drp-1}\Lambda(1-x_2)\nonumber\\
 &\quad+\dgp (1-x_1)^{\drp} x_{2}\Lambda(1-x_2) \Bigr).
\end{align}

We  call $(0,0;\epsilon)$ and $(1,\epsilon;\epsilon)$ trivial fixed points of the DE system \eqref{065644_25Jan14} 
Other fixed points can be written parametrically as $\bigl(x,x_{2}^\perp(x);\epsilon^\perp(x)\bigr)$ with $x\in (0,1) $. 
\begin{align*}
x_2^\perp(x)&=-\frac{1}{\beta}\ln\frac{1-x^{\frac{1}{\dlp-1}}}{(1-x)^{\drp-1}},\\
\epsilon^\perp(x)&=\frac{x_{2} (x) }{ (1-(1-x)^{\drp}\Lambda(1-x_{2}(x)) )^{\dgp-1}}.
\end{align*}
We call these fixed points non-trivial. 
\subsection{Potential Function of $\C$ at Fixed Points}
The DE system of $\C$ is given as follows. 
\begin{align*}
x_{1}^{(\ell+1)}&=(1-(1-x_{1})^{\dr-1} )^{\dl-1} \Lambda(1-(1-\epsilon)(1-x_{2})^{\dg-1}), \\
x_{2}^{(\ell+1)}&=(1-(1-x_{1})^{\dr-1} )^{\dl} \Lambda(1-(1-\epsilon)(1-x_{2})^{\dg-1})  .
\end{align*}
The potential function of the DE system is given as follows. 
\begin{align*}
&U(x_1,x_2,\epsilon)=\frac{\dg \dl}{\beta \dr}+1-\epsilon \nonumber \\
&-\Bigl(\frac{\dg \dl}{\beta \dr}(1-x_1)^{\dr}+(1-\epsilon)(1-x_2)^{\dg} \Bigr)\nonumber \\
&-\frac{\dg}{\beta}(1-(1-x_{1})^{\dr-1})^{\dl}\Lambda(1-(1-\epsilon)(1-x_{2})^{\dg-1}) \nonumber \\
&-\{\frac{\dg \dl}{\beta}x_1(1-x_1)^{\dr-1}+\dg(1-\epsilon)x_2(1-x_2)^{\dg-1}\Bigr).
\end{align*}
We call $(0,0;\epsilon)$ and $(1,1;\epsilon)$ trivial fixed point of the DE system. 
Other fixed points can be written parametrically as 
$\bigl(x,x_2(x);\epsilon(x)\bigr)$ with $x\in (0,1)$. 
\begin{align*}
x_2(x)&=x(1-(1-x)^{\dr-1}),\\
\epsilon(x)&=1+\frac{\ln\frac{x}{(1-(1-x)^{\dr-1})^{\dl-1}}}{\beta(1-x_{2}(x))^{\dg-1}}.
\end{align*}
We call these fixed points non-trivial. 

\subsection{Spatially-Coupled Precoded-Rateless Code $\widehat{\C}$ Achieves the Capacity}
In this section, we prove that SC $(\dl,\dr,\dg)$ precoded-rateless code $\hC$ achieves the capacity with $(\dl=2,\dr=3,\dg=3)$ and $(\dl=3,\dr=4,\dg=3)$. 
First we claim the following theorem. 
\begin{theorem}\label{MN-HA}
If the potential function of the DE system of $\Cp$ is positive at any non-trivial fixed point $(\xmn;\epsilon)$, i.e., 
$U^\perp(\xmn;\epsilon)>0$, 
then the potential threshold of the DE system of $\C$ is equal to the Shannon limit of $\C$. 
\end{theorem}
\noindent Proof:
Recall the definition of potential function. It is sufficient to show that 
\begin{align}
&U(\bm{1};\epsilon)=\es-\epsilon,\label{073543_25Jan14}\\
& \Uha(\xha(x_1);\epsilon(\xoh))>\ehs-\epsilon(\xoh). \label{205454_5Jan14}
\end{align}
Equation \eqref{073543_25Jan14} is obvious from 
\begin{align*}
\Uha(\bm{1};\epsilon)=\Umn(\bm{0};1-\epsilon)+\mu(\bm{1},1-\epsilon)-\nu(\bm{1};\epsilon)=\ehs -\epsilon
\end{align*}
followed from  \eqref{073849_25Jan14}. 
Next, we will show \eqref{205454_5Jan14}. 
Let $\bigl(\xha;\epsilon\bigr):=(\xoh,\xth;\epsilon)$ be a fixed point of $\C$. 
We see that $\bigl(\xmn:=(\xom,\xtm):=\f^\perp(\bm{1}-\xha;1-\epsilon(\xoh)),1-\epsilon\bigr)$ is a fixed point of $\Cp$. 
This maps trivial and non-trivial fixed points of $\C$ to trivial and non-trivial fixed points of $\Cp$, respectively. 
Therefore, from the assumption it follows that $\Umn(\f^\perp(\bm{1}-\xha;1-\epsilon(\xoh));1-\epsilon(\xoh))>0$. 
By using \eqref{073849_25Jan14} again, we have 
\begin{align}
\Uha(\xha;\epsilon(\xoh))&=\Umn(\f^\perp(\bm{1}-\xha;1-\epsilon(\xoh));1-\epsilon(\xoh))\\
& \hspace{5mm} +\mu(\bm{1},1-\epsilon(\xoh))-\nu(\bm{1},\epsilon(\xoh))\\
&> \mu(\bm{1},1-\epsilon(\xoh))-\nu(\bm{1},\epsilon(\xoh))\\
&=1-\frac{\dg}{\beta}(1-\frac{\dl}{\dr}) -\epsilon(\xoh) = \ehs -\epsilon(\xoh).
\end{align}
\qed

The SC $(\dl=2,\dr=3,\dg=3)$ precoded-rateless code $\hC$ has the smallest possible degree that satisfies the necessary condition of capacity achievability \cite{6620664}.
\begin{lemma}\label{lrg323}
Let $\Cp$ be the dual code of the $(\dl=2,\dr=3,\dg=3)$ precoded-rateless code. Let $\Umn$ be the potential function of $\Cp$. 
Then, for any non-trivial fixed point $(x,x_2^\perp(x),\epsilon^\perp(x))$,  it holds that
\begin{align}
\Umn(x, x_2^\perp(x_1),\epsilon^\perp(x_1))> 0 \mbox{ for }x\in (0,1).
\end{align}
\end{lemma}
Proof: To show $\Umn(x,x_2^\perp(x),\epsilon^\perp(x))>0$, it is sufficient to show that for $z\in(0,1)$, 
 \begin{align}
&\beta \left.\Umn(x,x_2^\perp(x),\epsilon^\perp(x))\right|_{z=\sqrt{x}}\label{012623_11Feb13}\\
&=:z(3-3z+z^{2})+(-3+2z+2z^{2}-2z^{3})\log(1+z)>0.
\end{align}
Using $-3+2z+2z^{2}-2z^{3}<0$ for $z\in(0,1)$ and giving an upper bound
\begin{align*}
 \log (1+z)< z-\frac{z^2}{2}+\frac{z^3}{3},\quad z\in(0,1)
\end{align*}
to  the second  term of \eqref{012623_11Feb13}, we have that  \eqref{012623_11Feb13} is greater than
\begin{align}
\frac{1}{6} z^2 (3+ 6z-14z^2+10z^3-4 z^4)=:\frac{1}{6} z^2 \phi(z).
\end{align}
Since $\phi''(z)=-4(7-15z+12z^{2})<0$ for $z\in (0,1)$and $\phi(0)=3>0,\phi(1)=1>0$, $\phi(z)>0$ for $z\in(0,1)$. This concludes the lemma. 
\qed

The SC $(\dl=2,\dr=3,\dg=3)$ precoded-rateless code $\hC$ has many bit nodes of degree two. This leads to high error floors. 
Next, let us consider SC $(\dl=3,\dr=4,\dg=3)$ precoded-rateless code. This has no bit nodes of degree two. 
\begin{lemma}\label{lrg343}
Let $\Cp$ be the dual code of the $(\dl=4,\dr=3,\dg=3)$ precoded-rateless code. Let $\Umn$ be the potential function of $\Cp$. 
Then, for any non-trivial fixed point $(x,x_2^\perp(x),\epsilon^\perp(x))$,  it holds that
\begin{align}
\Umn(x, x_2^\perp(x_1),\epsilon^\perp(x_1))> 0 \mbox{ for }x\in (0,1).
\end{align}
\end{lemma}
Proof: To show $\Umn(x,x_2^\perp(x),\epsilon^\perp(x))>0$, it is sufficient to show that for $z\in(0,1)$, 
\begin{align}\label{433U}
&\beta \left.\Umn(x,x_2^\perp(x),\epsilon^\perp(x))\right|_{z=x^{1/3}}\\
&=\frac{3z}{4}(4-8z^{2}+5z^{3})+(3-2z-2z^{3}+2z^{4})\log \Bigl( \frac{1-z}{(-1+z^{3})^{2}}\Bigr)\nonumber\label{233605_5Jan14}
\end{align}
By using $3-2z-2z^ {3} +2z^ {4}> 0$ and giving upper bounds 
\begin{align}
&\log(1+z+z^{2})\le z+\frac{z^{2}}{2}-\frac{2z^{3}}{3}+\frac{z^{4}}{4}+\frac{z^{5}}{5}\\
&\log(1-z)\le -z-\frac{z^{2}}{2}-\frac{z^{3}}{3}
\end{align}
for $z\in(0,1)$ to each term in \eqref{433U}, we have \eqref{433U} is greater than
\begin{align}
&\frac{1}{15} z^2 (30+55z^{2}-72z^{3}-212z^{4}+260z^{5} -12z^6-48z^7)\\
&=:\frac{1}{15} z^2 \psi(z).
\end{align}
Let $\psi_i(z),\ (i=0,\dotsc,9)$ be Sturm sequence \cite[p.~264]{gautschi2011numerical}of a polynomial $\psi(z)$. 
The number of sign changes of Sturm sequence at $z=0,1$ are both 4. 
From this, it follows that the number of roots of $\psi(z)=0$ is zero. 
In Table \ref{232505_5Jan14}, we listed the sign of Sturm sequence at at $z=0,1$. 
Since $\psi(0)=30,\psi(1)=1$, it follows that $\psi(z)>0$ for $z\in(0,1)$.
\qed

\begin{table}[t]
\begin{center}
\caption{The sing $\mathrm{sgn}[\psi_i(z)]$ of Sturm sequence $\{\psi_i(z)\}$ of $\psi(z)$.}
\label{232505_5Jan14}
  \begin{tabular}{c|cccccccccc}
  $i$ &0&1&2&3&4&5&6&7&8&9\\\hline
  $z=0$ &$+$&$0$&$-$&$-$&$+$&$+$&$+$&$-$&$-$&$+$\\
  $z=1$ &$+$&$-$&$-$&$-$&$-$&$+$&$+$&$-$&$-$&$+$
 \end{tabular}
\end{center}
\end{table} 

\begin{theorem}\label{3-4-3}
For sufficiently large $w$ and $L$, the spatially-coupled $(\dl,\dr,\dg)$ precoded-rateless code $\hC$ with coupling number $L$ and coupling width $w$
achieves the capacity of BEC for $(\dl,\dr,\dg)=(2,3,3), (3,4,3)$. 
\end{theorem}
Proof: 
The coding rate of $\hC$ with fixed coupling width $w$ converges to $\Rtotal$ in the limit of large $L$. 
Use  Lemma \ref{lrg323} and Lemma \ref{lrg343} as the condition of Theorem \ref{MN-HA}, then we see that 
the potential threshold of $(\dl,\dr,\dg)$ precoded-rateless code $C$ is equal to Shannon limit $\ehs:=1-\Rtotal=1-\frac{\dg}{\beta}\Bigl(1-\frac{\dl}{\dr}\Bigr)$.
Apply this result to Theorem \ref{thm:sc_theorem}, then we see that DE of $\hC$ converges to $\bm{0}$ if $\epsilon<\ehs. $
\qed

\begin{discussion}
In \cite{6620664}, we derived a necessary condition that $\hC$  achieves the capacity of BEC($\epsilon$) in the limit of $w$ and $L$ as follows. 
 \begin{align}
 \dg \ge \frac{\dr \ln (\dr-1)}{\dr-2}
 \end{align}
This is satisfied by $(\dl,\dr,\dg)=(2,3,3), (3,4,3)$. 
We observed that for many patterns of $(\dl,\dr,\dg)$ with no exception the potential function of $\hC$ at non-trivial fixed points is positive. 
From this observation, we believe that the necessary condition is actually also a sufficient condition for achieving capacity. 
\end{discussion}

\section{Conclusion and Future Work}
We proved that SC precoded-rateless codes achieve the capacity of BEC with bounded degree. 
The proof used duality property of potential functions over $\C$ and $\Cp$, 
since it is easier to prove that potential threshold is equal to Shannon threshold for $\Cp$ than $\C$. 

Future works include showing sufficient condition of capacity achievability on parameters $(\dl,\dr,\dg)$ and an extension to binary-input memoryless channels. 
 \section{Conclusion}
\bibliographystyle{IEEEtran}
\bibliography{IEEEabrv,../kenta_bib}
\end{document}